\documentclass{pasj00}

\usepackage{graphicx}
\usepackage{color}

\begin{document}
\SetRunningHead{Baba et. al.}{Interpretation of the $l-v$ features}
\Received{2010/3/29}
\Accepted{2010/9/14}

\title{On the Interpretation of the $l-v$ Features in the Milky Way Galaxy}

\author{Junichi \textsc{Baba}$^{1}$, 
       Takayuki \textsc{R. Saitoh}$^{2}$, and
       Keiichi \textsc{Wada}$^{3}$
}

\affil{$^1$ Center For Computational Astrophysics, National Astronomical
    Observatory of Japan, 2--21--1 Osawa, Mitaka-shi, Tokyo 181--8588.}

\affil{$^2$ Division of Theoretical Astronomy, National Astronomical
    Observatory of Japan, 2--21--1 Osawa, Mitaka-shi, Tokyo 181--8588.}

\affil{$^3$ Graduate School of Science and Engineering, Kagoshima University,
    1--21--30 Korimoto, Kagoshima, Kagoshima 890--8580.}

\email{baba.junichi@nao.ac.jp}

\maketitle

\begin{abstract}
We model the gas dynamics of barred galaxies using a three-dimensional, 
high-resolution, $N$-body+hydrodynamical simulation and apply it to the
Milky Way in an attempt to reproduce both the large-scale structure and
the clumpy morphology observed in Galactic H\emissiontype{I} and CO
$l-v$ diagrams. 
Owing to including the multi-phase interstellar medium, self-gravity,
star-formation and supernovae feedback, 
the clumpy morphology, as well as the large-scale features,
in observed $l-v$ diagrams are naturally reproduced.
We identify in our $l-v$ diagrams with a number of not only large-scale
peculiar features such as the `3-kpc arm', `135-km s$^{-1}$ arm' and 
`Connecting arm' but also clumpy features such as `Bania clumps', and
then link these features in a face-on view of our model. 
We give suggestions on the real structure of the Milky Way and on the
fate of gas clumps in the central region.
\end{abstract}

\section{Introduction}

In order to understand the structure and dynamics of the Milky Way
galaxy, the line-of-sight velocity of the interstellar medium (ISM) 
({\it e.g.}, H\emissiontype{I}, CO) is often used.
This is used to determine the radial mass distribution
\citep{SofueRubin2001} and the spatial distribution 
\citep{Oort1958, NakanishiSofue2003, NakanishiSofue2006}.
These studies had to adopt the fundamental assumption here is that the
ISM rotates axisymmetrically in a purely circular fashion on the
Galactic plane.  However, longitude-velocity ($l-v$) diagrams of
H\emissiontype{I} \citep{HartmannBurton1997} and CO \citep{Dame+2001} in
the inner Milky Way galaxy are inconsistent with the circular motions
\footnote{ See their Figure 1 and section 2 in
    \citet{Rodriguez-FernandezCombes2008} for a recent summary of
    outstanding peculiar features in the $l-v$ diagram of the ISM in the
    Milky Way galaxy.}.
Moreover, the central stellar bar and spiral arms result in a
non-axisymmetric gravitational potential.  Thereby, there is a large
difficulty in reconstructing the true mass distribution and spatial
distribution of the ISM from the line-of-sight velocities (
\cite{KodaWada2002,Gomez2006, Pohl+2008, Baba+2009} hereafter Paper I).

The peculiar features of the $l-v$ diagrams of H\emissiontype{I} and CO
in the Milky Way galaxy are widely accepted to be as non-circular flows
under the influence of the stellar bar and spiral arms
(see chapter 9 in \cite{BinneyMerrifield1998}).  \citet{Binney+1991}
modeled ISM motion in terms of closed stellar orbits ($x_1$ and $x_2$ 
orbits), and interpreted the peculiar features seen in $l-v$ diagrams of
the Galactic central region ($|l| < 10^\circ$) as non-circular motions
due to the Galactic bar. \citet{JenkinsBinney1994} tried to reproduce
the $l-v$ diagram of the central molecular zone (CMZ) by numerical
simulations using the so-called sticky-particle method.  
Many hydrodynamical simulations of the ISM have also been performed in order
to investigate the origin of the H\emissiontype{I} and CO $l-v$ diagrams
of the Galactic disk \citep{Wada+1994, EnglmaierGerhard1999, 
WeinerSellwood1999, Fux1999, Bissantz+2003, Rodriguez-FernandezCombes2008}. 
These previous numerical studies showed that the large-scale features of
the $l-v$ diagrams largely depend on the mass model of the bulge,
stellar bar, and spiral arms along with the pattern speed and the
location of the observer.

\citet{EnglmaierGerhard1999} performed hydrodynamic simulations of gas
flows in the gravitational potential of the near-infrared luminosity
distribution of the Milky Way \citep{Binney+1997}. 
Their best fit models qualitatively reproduced 
number of observed features in the $l-v$ diagrams.
\citet{Bissantz+2003} extended the model of \citet{EnglmaierGerhard1999} 
to include stellar spiral arms, and showed that gas flows in models with stellar
spiral arms match the observed $l-v$ diagram better than models
without stellar spiral arms.
\citet{Fux1999} performed three-dimensional $N$-body + hydrodynamical
simulations of the Milky Way, and gave a coherent interpretation of the main
features standing out from observed $l-v$ diagrams within the Galactic bar.
In particular, he showed that the tracers of the gas associated with the 
Milky Way's dust lanes can be reliably identified and the 3 kpc-arm appears 
as a gaseous stream rather than a density wave \citep{Fux2001}. 
He also argued that the density centre of the stellar bar wanders 
around the centre of mass and the resulting gas flow is asymmetric and 
non-stationary.

Advanced modeling of the Galactic disk provides further areas for
comparisons between the theoretical models and observations, and, in
addition, it can help us understand the true structure and dynamics of
the Galactic disk.
However, these previous numerical studies adopted a rather simple
modeling of the ISM and the stellar disk. 
First, an isothermal equation of state (EOS) with a velocity dispersion
of $\sim 10~{\rm km~s^{-1}}$ \citep{Wada+1994,EnglmaierGerhard1999,Fux1999,
WeinerSellwood1999,Bissantz+2003}, or a phenomenological model of the
ISM \citep{Rodriguez-FernandezCombes2008} were used. 
The isothermal EOS would be a relevant approximation to investigate the
global gas dynamics in galaxies, however the ISM is obviously not
isothermal.  It is important to include an energy equation with
appropriate cooling and heating terms in order to compare the numerical
results with the H\emissiontype{I} and CO observations 
({\it e.g.}, \cite{WadaSpaansKim2000}). The inhomogeneous nature of
the ISM should be taken into account to investigate the gas dynamics on
a local scale or in the central regions of galaxies
\citep{WadaKoda2001}. In fact, the observed $l-v$ diagrams ({\it e.g.}
\cite{Dame+2001}) show many `clumpy' sub-structures, which are not seen
in the previous computational $l-v$ diagrams.

Secondly, self-gravitational interactions of the ISM were often ignored 
\citep{EnglmaierGerhard1999,WeinerSellwood1999,Bissantz+2003}.
The self-gravity of gas can play a significant role in high-density
regions such as a central gaseous ring \citep{WadaHabe1992, Wada+1994,
Fukuda+2000}.  Furthermore, the self-gravity in gas along with the
thermal instabilities causes complicated, non-circular motions in the
multi-phase ISM \citep{Wada+2002}.  
A bar was modelled with a fixed potential undergoing a rigid
rotation, and stellar spiral arms were not considered
\citep{Wada+1994,EnglmaierGerhard1999,WeinerSellwood1999}.
In addition to the stellar bar, stellar spiral arms also result in
non-circular motions of the gas 
({\it e.g.}, \cite{Fujimoto1968, Roberts1969, Shu+1973}).  
\citet{Bissantz+2003} included stellar spiral arms, but the spiral arms
were assumed as rigid patterns.

In order to understand the origin of the both large-scale structures and
clumpy features in the observed $l-v$ diagrams, we present a
self-consistent high-resolution simulation of a disk galaxy,
which consists of a stellar disk and the multi-phase ISM in a static
dark-matter halo.  Star formation from cold, dense gas and energy
feedback from type II supernovae (SNe), which have not been included in
previous models, are taken into account.  We have simulated a model
galaxy with a smaller mass than the Milky Way galaxy where the circular
velocity at 8 kpc is 25 \% less than that in the Milky Way galaxy. This
enables us to ensure a high mass resolution for the ISM.  Despite this
difference, the $l-v$ diagram from our simulations can surprisingly
reproduces many features noted in observations, suggesting that our
method has potential to qualitatively reproduce features in the Milky
Way galaxy with a more massive model.  We are now preparing to run this
`Milky Way model' using ten times the number of particles 
({\it i.e.} $> 10^7$ particles) used in the work presented here.

In section 2, we describe our methodology and in section 3 we report the
numerical results. We present the qualitative comparison for these
structures of the Milky Way galaxy via $l-v$ diagrams in section 4.  In
section 5, we give a brief summary.

\section{Numerical Method and Model Setup}
\label{sec:method}

We used the $N$-body/hydrodynamic simulation code {\tt ASURA}
\citep{Saitoh+2008,Saitoh+2009} to solve the Newtonian equation of
motions and the equations of hydrodynamics using the standard smoothed
particle hydrodynamics (SPH) methods
\citep{Lucy1977,GingoldMonaghan1977,Monaghan1992} :
\begin{eqnarray}
\rho_i &=& 
    \sum_j^{N_{\rm nb}}m_jW(|\mathbf{x}_i-\mathbf{x}_j|,h),\\
\frac{d \mathbf{v}_i}{dt} &=& 
     - \sum_j^{N_{\rm nb}}m_j
     \left(\frac{p_i}{\rho_i^2} 
             + \frac{p_j}{\rho_j^2} + \Pi_{ij}\right)
     \nabla_iW(|\mathbf{x}_i-\mathbf{x}_j|,h)  \nonumber \\
     && + \mathbf{g}_i - \nabla\Phi_{\rm DM}(\mathbf{x}_i),\\
\frac{du_i}{dt} &=& 
     \sum_j^{N_{\rm nb}}m_j
     \left(\frac{p_i}{\rho_i^2}
             + \frac{1}{2}\Pi_{ij}\right)
     (\mathbf{v}_i-\mathbf{v}_j)
             \cdot\nabla_iW(|\mathbf{x}_i-\mathbf{x}_j|,h)  \nonumber \\
       &&  + \frac{\Gamma_i - \Lambda_i}{\rho_i},
\end{eqnarray}
where $m$, $\rho$, $p$, $u$, $\mathbf{v}$, $\mathbf{x}$, and $\Phi_{\rm DM}$
are the mass, density, pressure, specific internal energy, velocity, 
position of the gas, and the gravitational potential of the dark matter
halo, respectively. We assume an ideal gas EOS $p = (\gamma-1)\rho u$, 
with $\gamma =5/3$.
$W(x,h)$ and $h$ are the SPH smoothing kernel and the smoothing length, 
respectively, and $h$ is allowed to vary both in space and time with
the constraint that the typical number of neighbours for each particle
is $N_{\rm nb} = 32\pm2$. 
The artificial viscosity term $\Pi_{\rm ij} $ \citep{Monaghan1997} and
the correction term to avoid large entropy generation in pure shear
flows \citep{Balsara1995} are used.
Radiative cooling of the gas, $\Lambda$, was solved assuming an
optically thin cooling function with solar metallicity which covered a
wide range of temperature, $20$ K through $10^8$ K
\citep{WadaNorman2001}.  
Heating due to far-ultraviolet radiation (FUV) and energy feedback from
SNe, $\Gamma = \Gamma_{\rm FUV} + \Gamma_{\rm SN}$, was also included.
We assume a uniform FUV field with matches that observed in the solar
neighborhood: 
\begin{equation}
 \Gamma_{\rm FUV} = 10^{-24}\epsilon G_0 n_{\rm H}~
 {\rm [erg~s^{-1} cm^{-3}]}
\end{equation}
\citep{Wolfire+1995}, where $\epsilon$ and $G_0$ are the heating
efficiency ($\epsilon = 0.05$) and incident FUV normalized to the solar
neighborhood value ($G_0 = 1$), respectively.  The self-gravity of stars
and SPH particles, $\mathbf{g}$, is calculated by the Tree with GRAPE
method \citep{Makino1991}. We here used a software emulator of GRAPE,
Phantom-GRAPE (Nitadori et al. in preparation).  

Models for the star formation and the SN feedback were the same as those
in Paper I and \citet{Saitoh+2008}.
We adopted the single stellar population approximation, with the 
Salpeter initial mass function \citep{Salpeter1955} and the mass range 
of $0.1-100~{\rm M}_{\odot}$. 
If an SPH particle satisfies the criteria 
(1) $n_{\rm H} > 100~{\rm cm^{-3}}$, (2) $T < 100~{\rm K}$, 
and (3) $\nabla\cdot\mathbf{v}<0$, the SPH particle creates star particles 
following the Schmidt law \citep{Schmidt1959}, with a local star
formation efficiency, $C_{\ast}=0.033$, in a probabilistic manner
\citep{TaskerBryan2006,Saitoh+2008,Saitoh+2009,
    TaskerBryan2008,RobertsonKravtsov2008}.
The local star formation efficiency $C_{\ast}$ is an unknown
parameter.
\citet{Saitoh+2008} however showed that global (galactic) star
formation rate is not directly proportional to the local star formation
efficiency, $C_{\ast}$, 
but mainly controlled by gas mass in the high density regions, 
which is statistically related to the global evolution of the ISM
 (for details, see Section 5.2 in \cite{Saitoh+2008}). 
We treated Type II SNe feedback
as injecting the thermal energy from SNe to the neighbour SPH particles.
We here assumed that stars with $> 8~M_{\rm \odot}$ experience Type II
SNe and each SN release an energy with a canonical value of $10^{51}$
ergs.
It was suggested that if the feedback energies are lower than the
canonical value, it is hard to reproduce hot, diffuse halos around disk galaxies
\citep{TaskerBryan2006,TaskerBryan2008}.

We follow the same method presented in Paper I: we first embed
a pure $N$-body stellar disk with an exponential profile within a static
dark-matter (DM) halo potential whose density profile follows the
Navarro-Frenk-White profile \citep{Navarro+1997}.  Using Hernquist's
method \citep{Hernquist1993}, we generate the initial conditions for the
stellar disk which is then allowed to evolve for 2 Gyr, resulting in the
formation of a stellar bar and multi-arms.  At this point, we added a
gaseous component to the stellar disk. This combined fixed dark-matter
potential embedded stellar disk + gas disk is used as the initial
condition for the simulation presented in this paper. The model
parameters of the dark matter halo, stellar disk, and gas disk are
summarized in Table \ref{tbl:model}. 
The circular velocity curve is shown in Figure \ref{fig:angfreq}a.

The total numbers of stars and gas particles are $3\times 10^6$ and
$10^6$ respectively, and particle masses are $11000~\Mo$ and 
$3200~\Mo$. This gas particle mass allows us to resolve
the gravitational fragmentation of dense clumps of gas down to $\sim
10^5~\Mo$ \citep{Saitoh+2008}. In addition, a gravitational softening
length of $10~{\rm pc}$ in association with a typical smoothing length
in dense regions of a few tens of pc means that structures larger than a
few tens of pc are well resolved.

\begin{table}[htdp]
\caption{Model parameters for each mass component (dark halo, stellar
        disk, and gas disk)}
\begin{center}
\begin{tabular}{ lll }
\hline
\hline 
 Component	&	Parameter	& Value	\\
\hline
Dark Halo			& Mass	&	$6.3\times 10^{11}$ M$_{\odot}$\\
								& Radius 	&  $122$ kpc\\
								& Concentration 	& $5.0$ \\
\hline
Initial Stellar Disk	& Mass	&  $3.2\times 10^{10}$ M$_{\odot}$\\
								& Scale Length 	& $3.0$ kpc \\
								& Scale Height	& $0.3$ kpc\\
\hline
Initial Gas Disk	& Mass	& $3.2\times 10^{9}$ M$_{\odot}$\\
								& Scale Length	& $6.0$ kpc\\
								& Scale Height	& $0.2$ kpc \\
\hline
\end{tabular}
\end{center}
\label{tbl:model}
\end{table}%

\section{Results}

\subsection{Three-dimensional structures}
\begin{figure*}
\begin{center}
\includegraphics[width=0.90\textwidth]{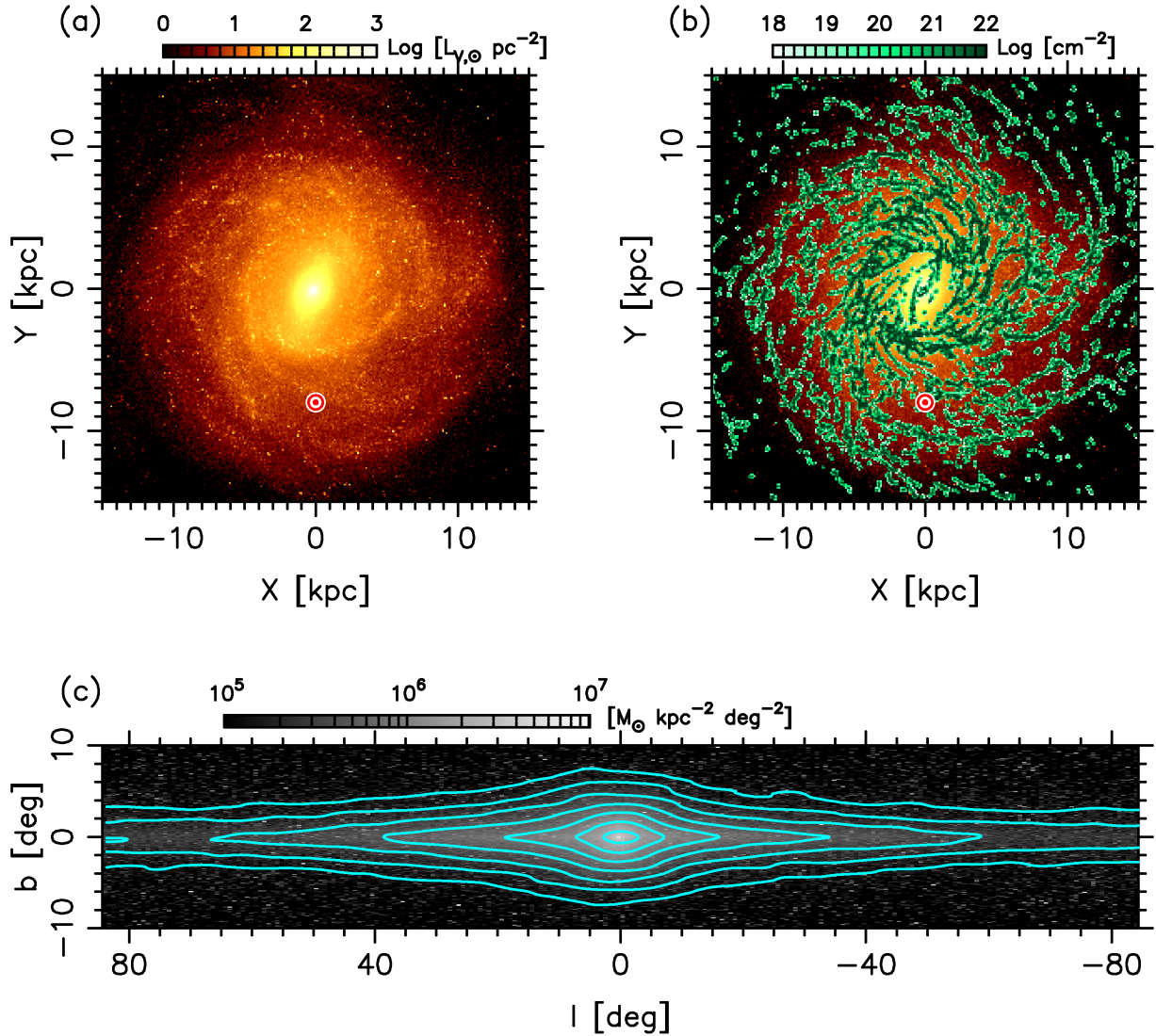}
\caption{
    (a): V-band image of the stellar disk (orange) at $t=1.24~{\rm Gyr}$.  
    A major axis of the stellar bar inclines toward $25^\circ$ from a $y$-axis. 
    The galactic rotation is clock-wise.
    (b): Overlaid of the cold gases (green, $T < 100~{\rm K}$) on the
        panel (a).
    (c): Distribution of stellar surface density on the $l-b$ plane. 
    The observer is located at the red point ($0~{\rm kpc},-8~{\rm
            kpc}$) in the panels (a) and (b).}
\label{fig:snapshot}
\end{center}
\end{figure*}

In Figure \ref{fig:snapshot}a, we present a face-on view of the stars
at $t=1.24~{\rm Gyr}$ where $t=0$ is the time when the gas component is
added to the pre-evolved stellar disk.  
By this time, we see a stellar bar with a semi-major axis of $\sim 3-4$
kpc. 
In R = $\sim 5$-$10$ kpc, two-armed stellar
spiral dominates, and outside $\sim 10$ kpc multi-armed spirals appear
(see also Figure 2 and 4 in Paper I). 
The Fourier amplitude of the dominant mode is $\sim 0.1-0.2$, which is 
consistent with observational values of external disk galaxies
\citep{RixZaritsky1995,Zibetti+2009}.

Figure \ref{fig:snapshot}b present a face-on view of the cold 
gas ($T < 100~{\rm K}$) overlaid on the stellar disk.
Along the bar a typical offset ridge structure of cold gas (dust lane)
is formed and enhanced by the bar potential. 
There are many gas arms in the disk, but contrast to previous isothermal
simulations ({\it e.g.}, \cite{Fux1999}), these arms are not smooth.
There are also many substructures, such as clumps and filaments, in the
inter-arm regions.  Hence, it is not easy to trace any ``single arm''
from the central region to the outer disk, even for major spiral arms 
(see also Figure \ref{fig:faceonlv}c).
Typical length and width of the gas filaments are $<500$ pc and $\sim
100$ pc, respectively. 
These local structures, i.e., gas clumps and filaments, are originated
from their self-gravity, whereas large-scale structures, i.e., gaseous
spiral arms, are driven by the stellar bar and spiral arms.

Figure \ref{fig:snapshot}c shows the surface density of the stars on the
$(l,b)$-plane. The observer is assumed to be at $(R_0,\phi_{\rm b}) =
(8~{\rm kpc}, 25^\circ)$, where $R_0$ and $\phi_{\rm b}$ are the
galacto-centric distance and position angle relative to the major
axis of the bar, respectively (positive sign means for the anti-galactic 
rotation).  
Since many previous studies suggest that the orientation of the
major axis of the Galactic bar relative to us typically ranges from
$15^\circ$ to $35^\circ$ ({\it e.g.}, 
\cite{Binney+1997,Fux1999,EnglmaierGerhard1999,Bissantz+2003}; 
for a review, \cite{Merrifield2004}), 
we adopt a fiducial value of $\phi_{\rm b} = 25^\circ$.  Due to our
choice of $\phi_{\rm b} = 25^\circ$ 
we observe an asymmetry in disk and bulge densities in longitude. 
This asymmetry agrees with near-infrared observational studies by
{\it COBE/DIRBE} \citep{Weiland+1994,Dwek+1995,Binney+1997}.

\begin{figure}[htbp]
\begin{center}
\includegraphics[width=0.40\textwidth]{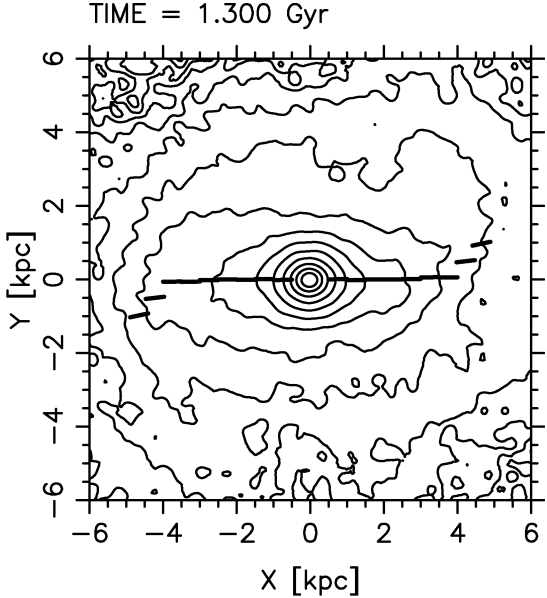}
\caption{Surface density contours of the stellar disk at $t= 1.3$ Gyr
    are spaced by a constant interval of 0.2 dex. Directions of the
        principal axis from $R=0.5$ kpc to $5.0$ kpc every 0.5 kpc are
        indicated by thick lines. 
        The galactic rotation is clock-wise.}
\label{fig:contour}
\end{center}
\end{figure}

\begin{figure}[htbp]
\begin{center}
\includegraphics[width=0.40\textwidth]{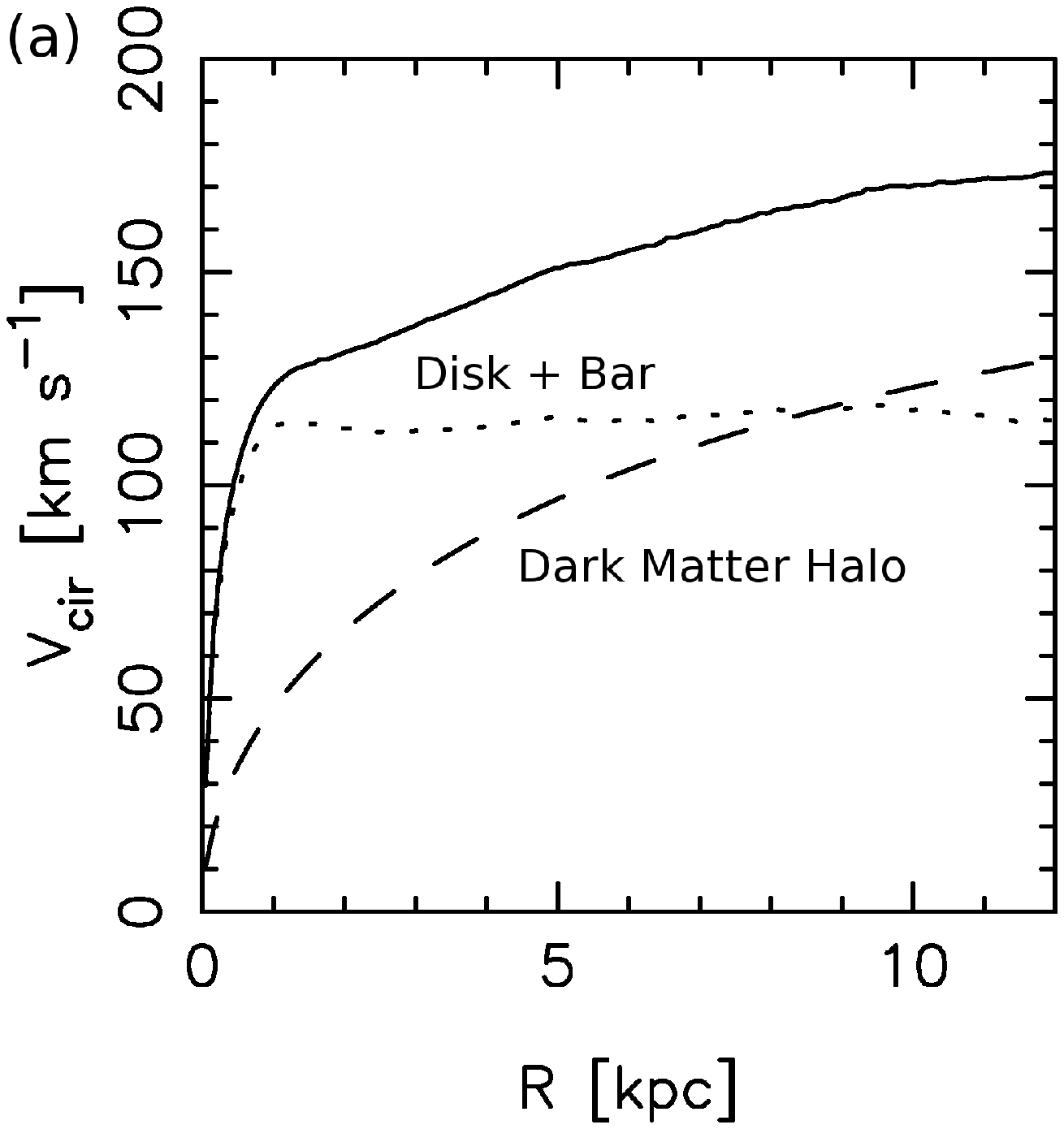}
\includegraphics[width=0.40\textwidth]{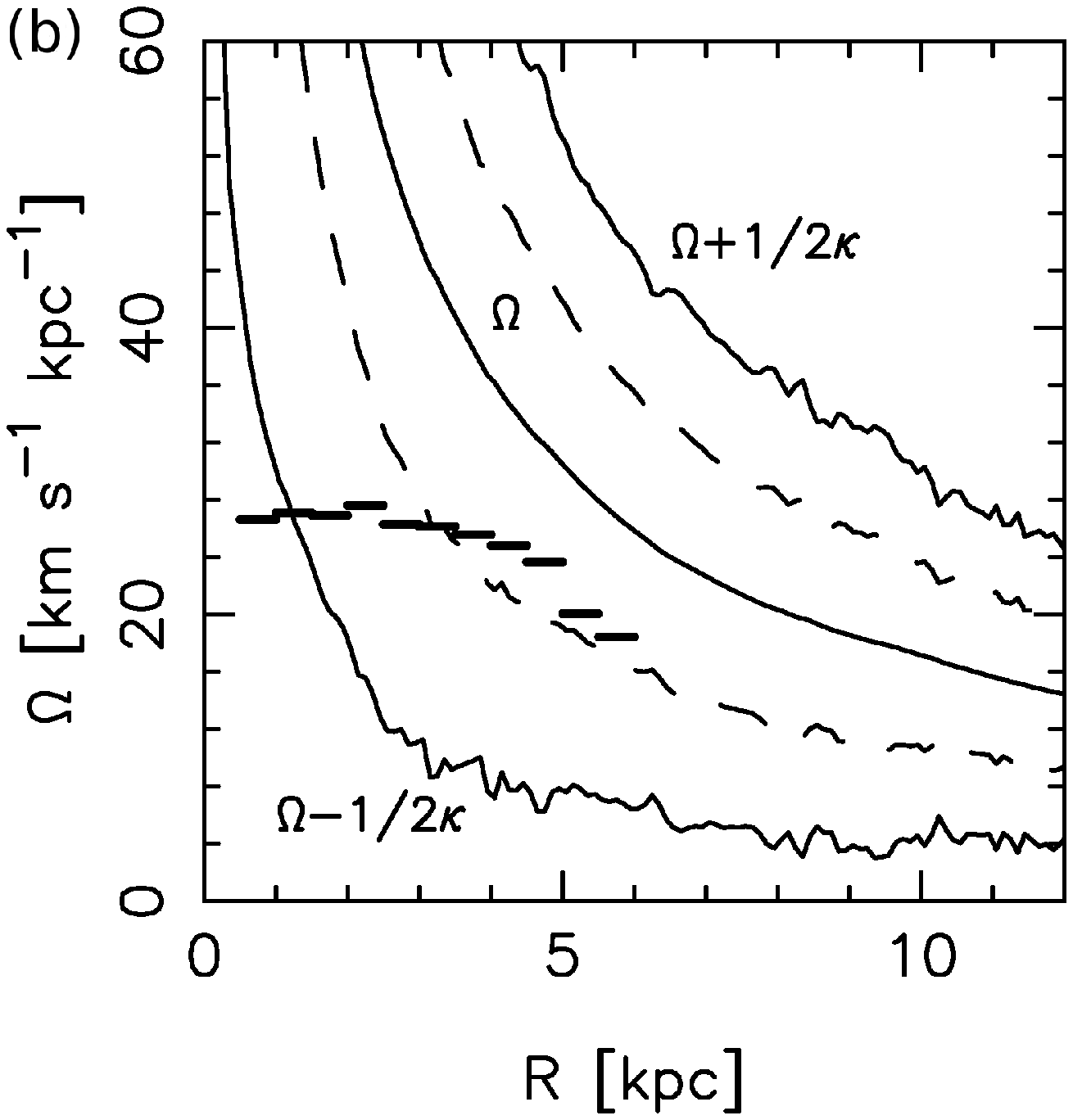}
\caption{(a) Circular velocities of the galaxy model. The circular velocities
    (solid curve) are derived from the azimuthally averaged potential.
    The contribution from dark matter halo is shown by a dashed curve.
    (b) Angular frequencies as a function of radius. Rotation angular
    ($\Omega$) and epicyclic ($\kappa$) frequencies are derived from the
    azimuthally averaged potential.  The dashed curves indicate
    $\Omega-1/4\kappa$.  Thick horizontal lines are pattern speeds of
    the principal axes presented in Figure \ref{fig:contour}.}
\label{fig:angfreq}
\end{center}
\end{figure}

Surface density contours of the stellar disk are shown in Figure
\ref{fig:contour}. The principal axis of the surface density
distribution for a given annulus of width $\Delta R = 0.5$ kpc are
indicated by thick lines. Within a radius $R < 4$ kpc,  they well align
in a same axis however, the outer annuli show a different distribution
axis.  Figure \ref{fig:angfreq}b shows pattern speeds of the principal
axis.  The pattern speeds for $R <4$ kpc are more or less the same 
at $\sim  27$ km s$^{-1}$ kpc$^{-1}$, however further out the pattern
speed declines. Therefore, we determine that the semi-major axis of the
stellar bar ($R_{\rm b}$) is $\sim 4 $ kpc with a face-on axis ratio of
$\sim 0.5$ and the bar rotates as a rigid-body with $\Omega_{\rm b}
\sim 27$ km s$^{-1}$ kpc$^{-1}$.  Comparing $\Omega_{\rm b}$ and the
angular frequency of the galaxy ($\Omega$), we infer the location of the
co-rotation radius ($R_{\rm CR}$) be around $R = 5$ kpc, whose value is
larger than $R_{\rm b}$.  
This is consistent with the theoretical studies 
({\it e.g.}, \cite{Contopoulos1980}) where a self-consistent bar is
required to have $R_{\rm CR}/R_{\rm b} > 1$, as well as observations 
of other barred galaxies ({\it e.g.}, \cite{Aguerri+2003}).
The inner and outer Lindblad resonances are located around $R = 1$ kpc 
and $10$ kpc, respectively.  
We note that the values of these bar parameters do not significantly change 
during the simulation ($t = 2$ Gyr).
The inner Lindblad resonance (ILR) locates at $\sim 1$ kpc in this
model, which is quite different from some previous models ({\it e.g.},
\cite{Binney+1991,Bissantz+2003}) that place the ILR at $\sim 100$ pc.
This difference may be caused by differences of the mass model in the
central region between our model and the Milky Way (see section 4). 

\subsection{The $l-v$ features}
\begin{figure*}[htbp]
\begin{center}
\includegraphics[width=.80\textwidth]{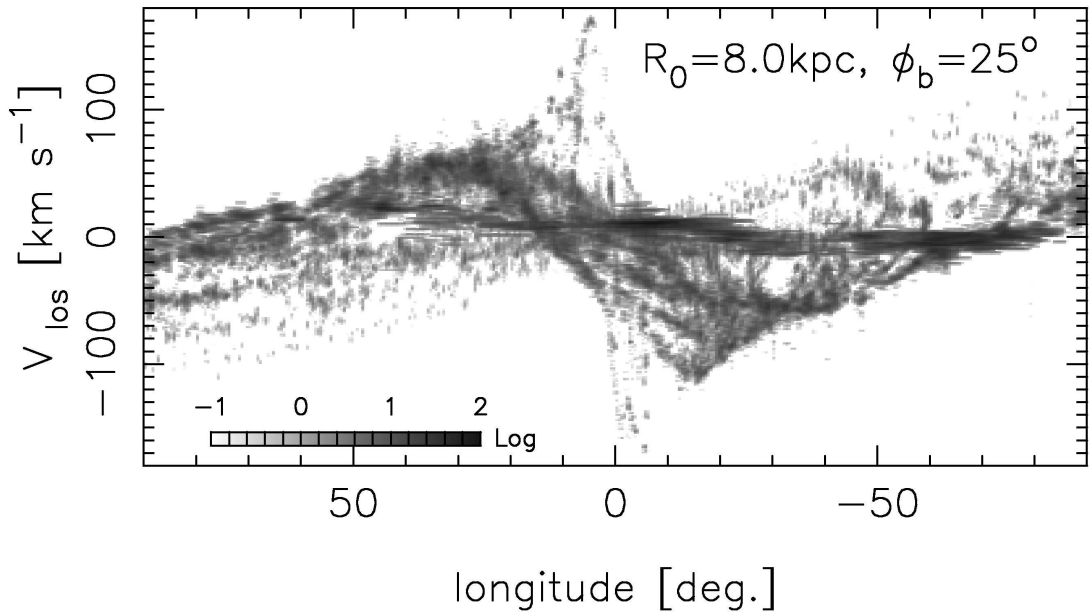}
\end{center}
\caption{Synthetic $l-v$ diagram in the range $|b| < 1^\circ$ derived
    from the cold gas ($T<100$K) distribution given in Figure
        \ref{fig:snapshot}a ($t=1.24$ Gyr).  The observer is located at
        the red mark point indicated in Figure \ref{fig:snapshot}, and
        has pure circular motion. The contribution of each SPH particle
        to the synthetic $l-v$ diagrams is weighted by its inverse
        squared distance relative to the observer in order to mimic the
        flux decline of point source. The gases with $d <500$ pc are not
        displayed.} 
\label{fig:pv}
\end{figure*}

\begin{figure}[htbp]
\begin{center}
\includegraphics[width=.40\textwidth]{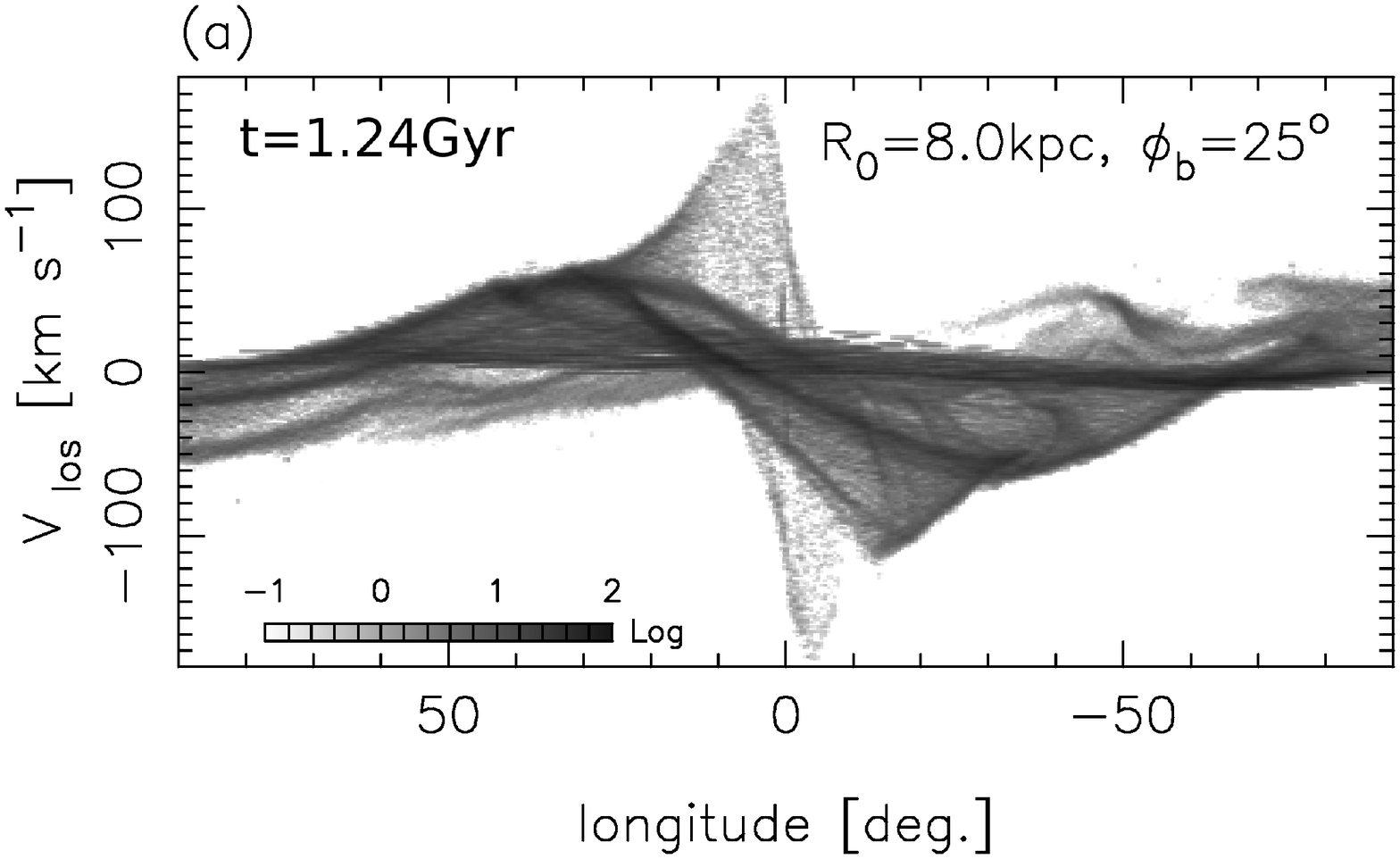}
\includegraphics[width=.40\textwidth]{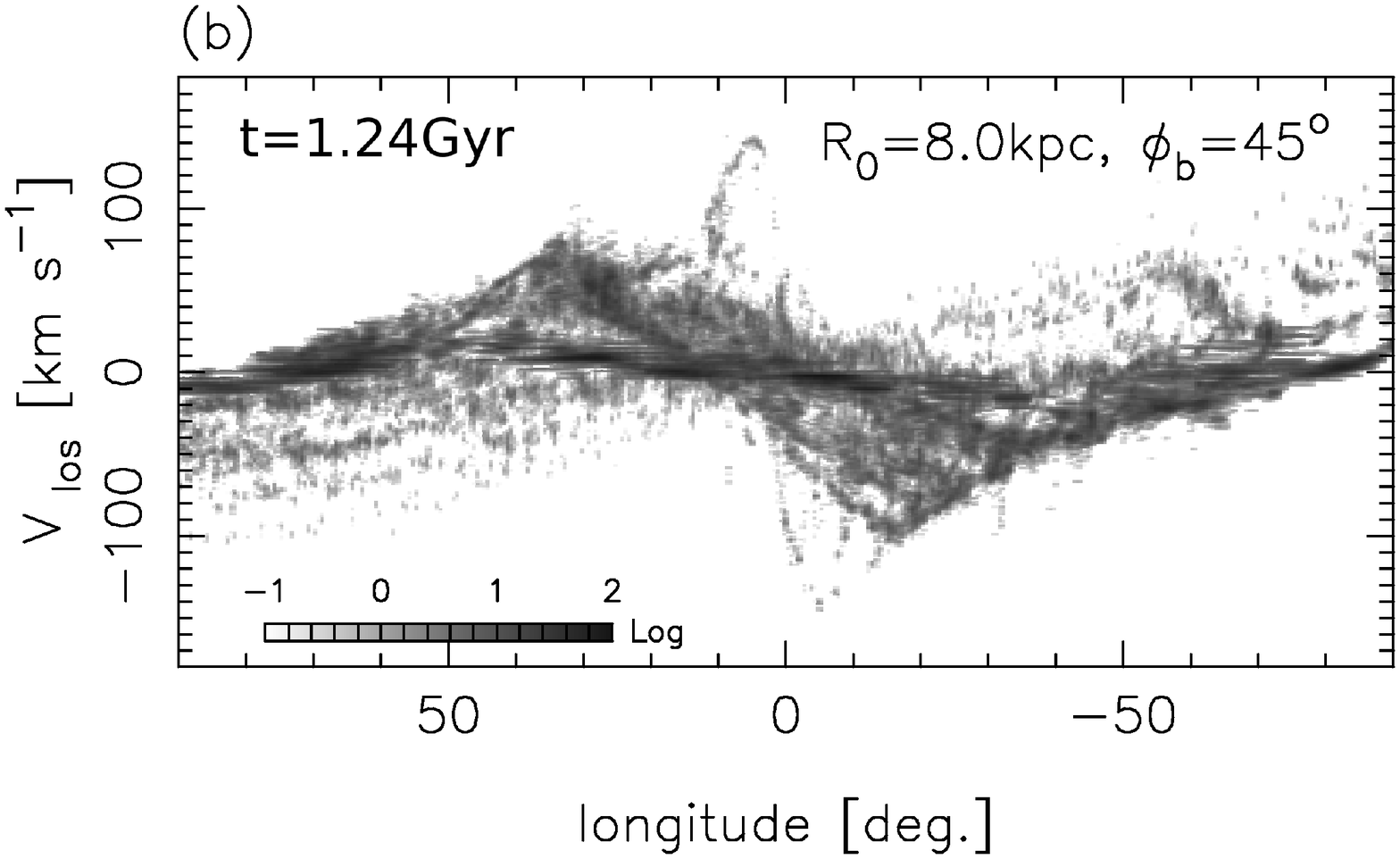}
\includegraphics[width=.40\textwidth]{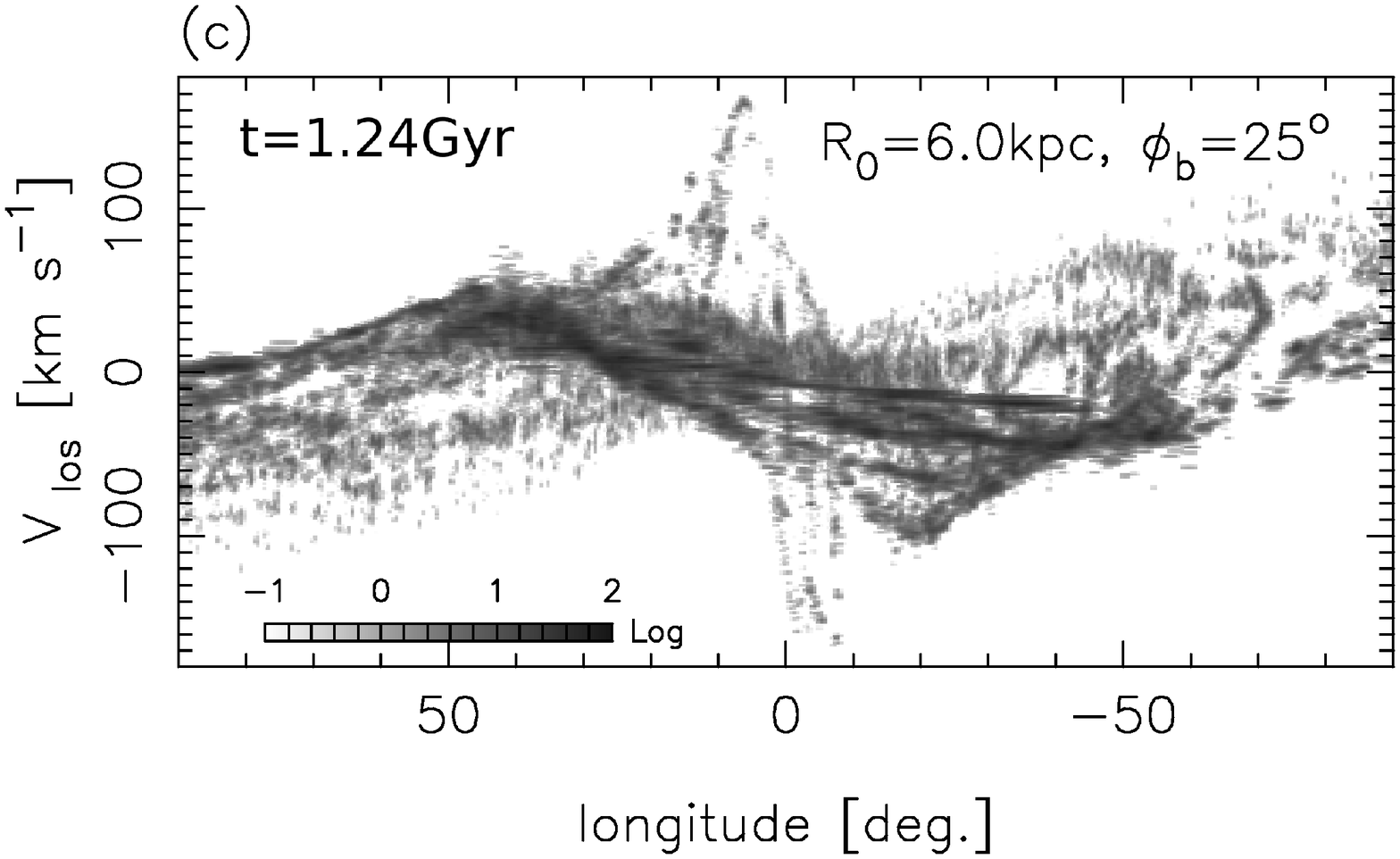}
\includegraphics[width=.40\textwidth]{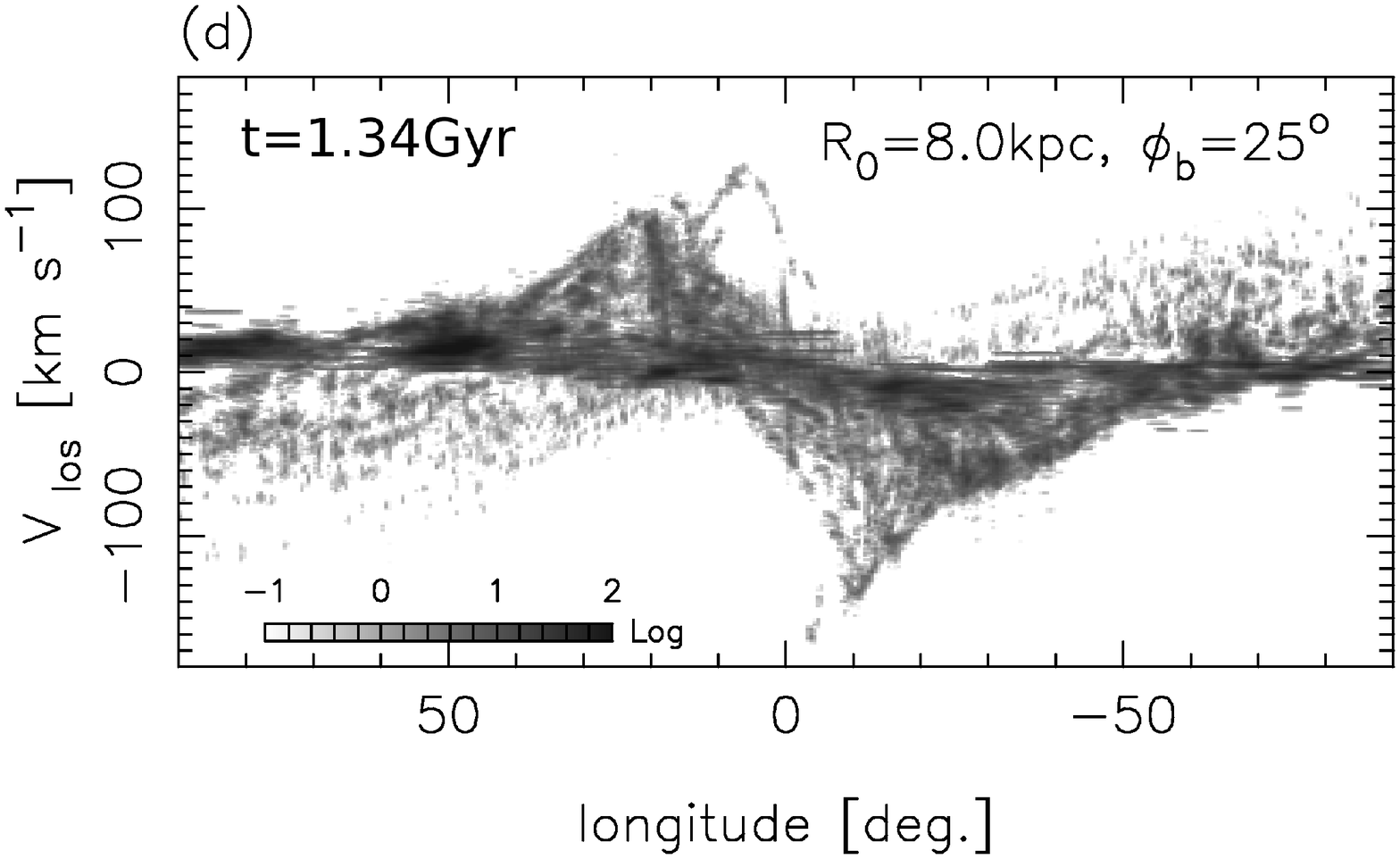}
\end{center}
\caption{Same as Figure \ref{fig:pv}, but for 
 (a) the isothermal gas with $T=10^4$K, (b) $\phi_{\rm b} = 45^\circ$,
 (c) $R_0=6$kpc, and (d) $t=1.34$ Gyr.}
\label{fig:pvcompare}
\end{figure}

Figure \ref{fig:pv} shows a synthetic $l-v$ diagram derived from the
cold gas distribution given in Figure \ref{fig:snapshot}a.  The diagram
shows not only the peculiar features but also the terminal velocity
tangent points observed in the H\emissiontype{I} and CO $l-v$ diagrams.  
Furthermore, our synthetic $l-v$ diagram shows a clumpy morphology whose
features posses a large velocity width 
(typically $\sim 10-20$ km s$^{-1}$), in agreement with CO $l-v$
diagrams \citep{Dame+2001}.  These clumpy features have not been
reproduced in the previous studies (\cite{Wada+1994,EnglmaierGerhard1999,
Bissantz+2003,Fux1999,Rodriguez-FernandezCombes2008}).  
We also show the $l-v$ diagram from a simulation with an isothermal EOS
\footnote{We replaced the gas in the multi-phase run into the isothermal 
gas with $T=10^4$K at $t=1.0$ Gyr, and then let it evolve for $240$ Myr.
Thus the stars distribute almost same as in the multi-phase run.} in 
Figure \ref{fig:pvcompare}a. Here this model includes the self-gravity
of the ISM. Unlike the $l-v$ diagram of the run with the multi-phase
ISM, this $l-v$ diagram shows a smooth rather than clumpy morphology.  
This result indicates that the appropriate treatment of the gas physics
in ISM, especially cooling down to $T \ll 10^4$ K, is essential to
understand clumpy morphology formed in the observed $l-v$ diagrams.

\begin{figure*}[htbp]
\begin{center}
\includegraphics[width=.80\textwidth]{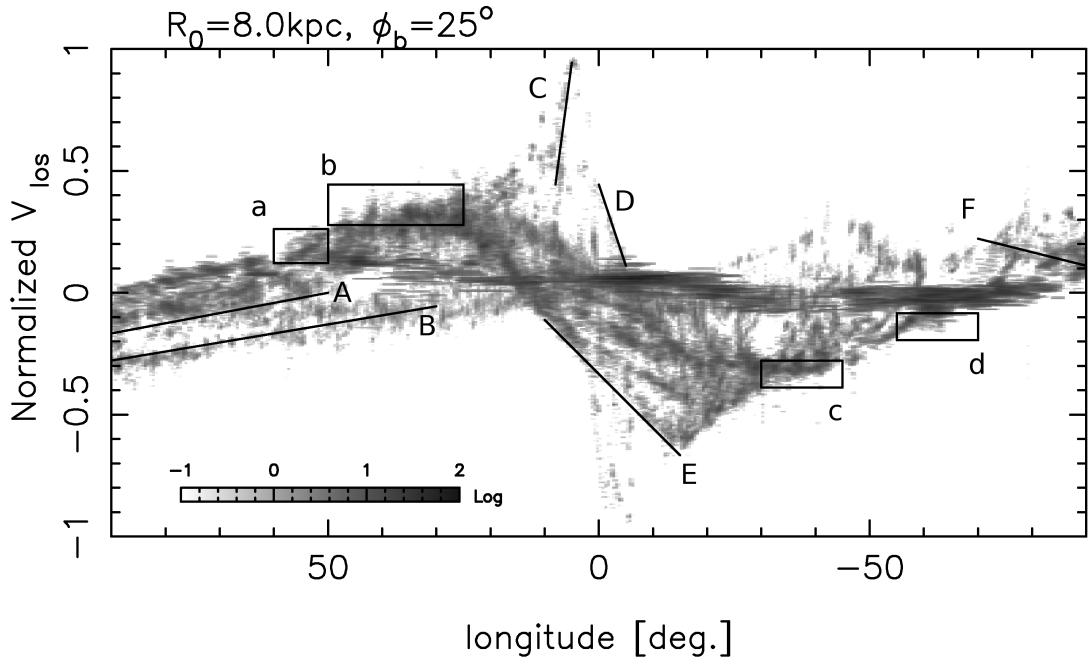}
\end{center}
\caption{Same as Figure \ref{fig:pv}, but for the line-of-sight velocity
    normalised to $180~{\rm km~s^{-1}}$. Schematic tracers of peculiar
    features (lines) and terminal velocity tangency (squares) are
    shown. See the text for the labeled lines and boxes.}
\label{fig:pvschematic}
\end{figure*}

The features in $l-v$ diagrams depend on both the location of the
observer within the galactic plane and the time.  To illustrate this, we
show $l-v$ diagrams for $(R_0,\phi_{\rm b})=(8~{\rm kpc}, 45^\circ)$ and
$(6~{\rm kpc}, 25^\circ)$ in Figure \ref{fig:pvcompare}b and
\ref{fig:pvcompare}c, respectively. If we observe the nuclear region of
the simulation from $\phi_{\rm b} = 45^\circ$, the velocity of the core
is lower than that for the case with $\phi_{\rm b} =25^\circ$.  The
diagonal features, such as so-called `3-kpc arm' 
(labeled by E in Figure \ref{fig:pvschematic}) seen in the CO $l-v$
diagrams, also become shallower. 
The position of the `Carina arm' (labeled by F in Figure \ref{fig:pvschematic}) 
around $l\sim -50^\circ-60^\circ$ and $v \sim 30~{\rm km~s^{-1}}$ also shifts
to larger $l$. Changing the radial position of the observer to $R = 6$
kpc alters the $l-v$ diagrams such that we see a larger number of
high-velocity components in the regions of $l < 0, v > 0$ or $l > 0, v <
0$. The $l-v$ diagram at $t=1.34$ Gyr, which is $100$ Myr after the
snapshot used for producing Figure \ref{fig:pv}, is shown in Figure
\ref{fig:pvcompare}d. One can notice that the non-axisymmetric features
change on this time-scale. This reflects the fact that the stellar and
gaseous spiral arms are not stationary, but time-dependent phenomena.

\section{A comparison to real structures in the Milky Way}
\label{sec:discussion}

\subsection{Large-Scale Structures}
A qualitative comparison between the numerical results and observations
of the Milky Way is given in Figure \ref{fig:pvschematic}. 
We identify and trace schematically (solid lines) such peculiar features
as the Perseus (A), Outer (B), connecting (C), 135-km s$^{-1}$ (D), 
3-kpc (E), and Carina (F) arms. 
The terminal velocity tangent points such as the Sagittarius (a), Scutum
(b), Norma (c), and Centaurus (d) tangent points, are also shown
schematically by squares. Although the Perseus arm, Sagittarius tangent
point, and central molecular zone (CMZ) are unclear in our model, 
other structures are clearly evident.  

In order to see which features in the $l-v$ diagram correspond to
structures in the real space, we marked several major features with the
same color in the face-on map (Figure \ref{fig:faceonlv}a, b).  
This comparison gives us insight to infer the real morphology of spiral
arms in the Milky Way.  For example, a prominent feature known as the
``molecular ring'' observed in CO $l-v$ diagrams is not necessarily
correspond to a ``ring'' in real space, but be a part of nearby spiral
arm (such as the the one colored by light blue).
\citet{NakanishiSofue2006} suggest that the molecular ring results from
a combination of the inner part of the Sagittarius-Carina arm and the
Scutum-Crux arm.  The ``3-kpc arm''(E) corresponds to an inner spiral
arm (the one colored by green).  
It has been proposed that this arm arises from a lateral arm surrounding
the bar (Figure 16 in \cite{Fux1999}), or a small arm starting from the
bar-end (Figures 13 and 15 in \cite{Bissantz+2003}). Following the
interpretation by \citet{Bissantz+2003}, the 3-kpc arm 
would correspond to the gaseous arm colored in blue in Figures \ref{fig:faceonlv}a,b.
Therefore, our result supports to the interpretation by \citet{Fux1999}.  
The near part of the offset ridge is the one colored red, which could be
due to a part of the ``connecting arm''(C). 
Therefore, the clumpy nature in this arm means that the offset ridge
of the Milky Way galaxy can consist of some gas clumps.  Contrary to
\citet{Fux1999}, ``$135$-km s$^{-1}$ arm'' (D) is a part of the far side
of the bar end in our model.  
We note, however, that above arguments on this comparison between
corresponding structures in our model and observations is qualitative, 
because our current ``barred'' galaxy is somewhat smaller than
the Milky Way resulting in a lower rotational velocity at $8~{\rm kpc}$
of $163~{\rm km~s^{-1}}$ compared to $\sim 220~{\rm km~s^{-1}}$.

The cold gas in the galactic center ($< 200$ pc, the one colored by 
purple) corresponds to the feature known as CMZ in the observed CO $l-v$
diagram.  However, it is not clearly seen in our $l-v$ diagram model in
comparison to that in the Milky Way galaxy. 
On the mass modeling, this may be attributable to differences in the bar
parameters between our model and the Galactic bar, but also to absence
of a central bulge.
It is known that bar parameters, such as the pattern speed and axis
ratio, can affect a gas inflow rate and shape of dust lanes, thereby
influence formation of the CMZ 
\citep{Athanassoula1992, EnglmaierGerhard1997,PatsisAthanassoula2000}.
According to these previous studies, the Galactic bar may be rounder or
more concentrated than the bar obtained in our simulation. Another
discrepancy between our model and the Milky Way is that stellar number
counts show evidence for the existence of an inner bar in the Milky
Way, which is much smaller than the outer bar (or triaxial bulge) with  
a semi-major axis of $3.5$ kpc \citep{Alard2001, Nishiyama+2005}
\footnote{
There is a discussion on the existence of the inner bar in the Milky
Way. At the central region, the magnitude differences in the star counts
on both sides are less than $0.1$ mag, so effects from asymmetric dust
absorption will cause quite some uncertainly in the interpretation of
the star counts. However, observations of external barred galaxies shows
that many barred galaxies have inner bars ({\it e.g.}, \cite{Erwin2004}). 
}.
However, our model does not have nested stellar bars.
\citet{Rodriguez-FernandezCombes2008} suggest that the parallelogram
shape of the $l-v$ diagram of the CMZ is due to the influence of this
inner bar (also known as nuclear bar or secondary bar).  
The central bulge, which is not included in the initial condition of 
the present model, may also affect the dynamics of the central region. 
More detailed comparison between the observed kinematics of 
the gas and models in the central several degrees requires 
high numerical resolution in the central part in numerical models. 
Effects of the central bulge and bar properties on 
the kinematics will be discussed in the high resolution models elsewhere.

In terms of realistic modelling heating physics of the ISM, 
we need to model the spatial dependence of the incident FUV, 
which has been assumed to be constant ($G_0=1.0$) everywhere in this paper 
(see equation 4). \citet{GerritsenIcke1997} calculated the incident FUV
by summing the FUV flux from all stars according to their ages. The
resultant radiation field declines outside the stellar disk with radius
as $1/R^2$.  This means that the incident FUV is stronger in the central
region than that in the solar neighborhood, implying that if we were to
consider the spatial dependence of the incident FUV in our model, for
example $G_0 \propto 1/R^2$, then the consumption of gas by star
formation in the region of the CMZ might be suppressed, yielding a
stronger CMZ present in our $l-v$ diagram.

\subsection{Clumpy Morphology}
\begin{figure*}[htbp]
\begin{center}
\includegraphics[width=.90\textwidth]{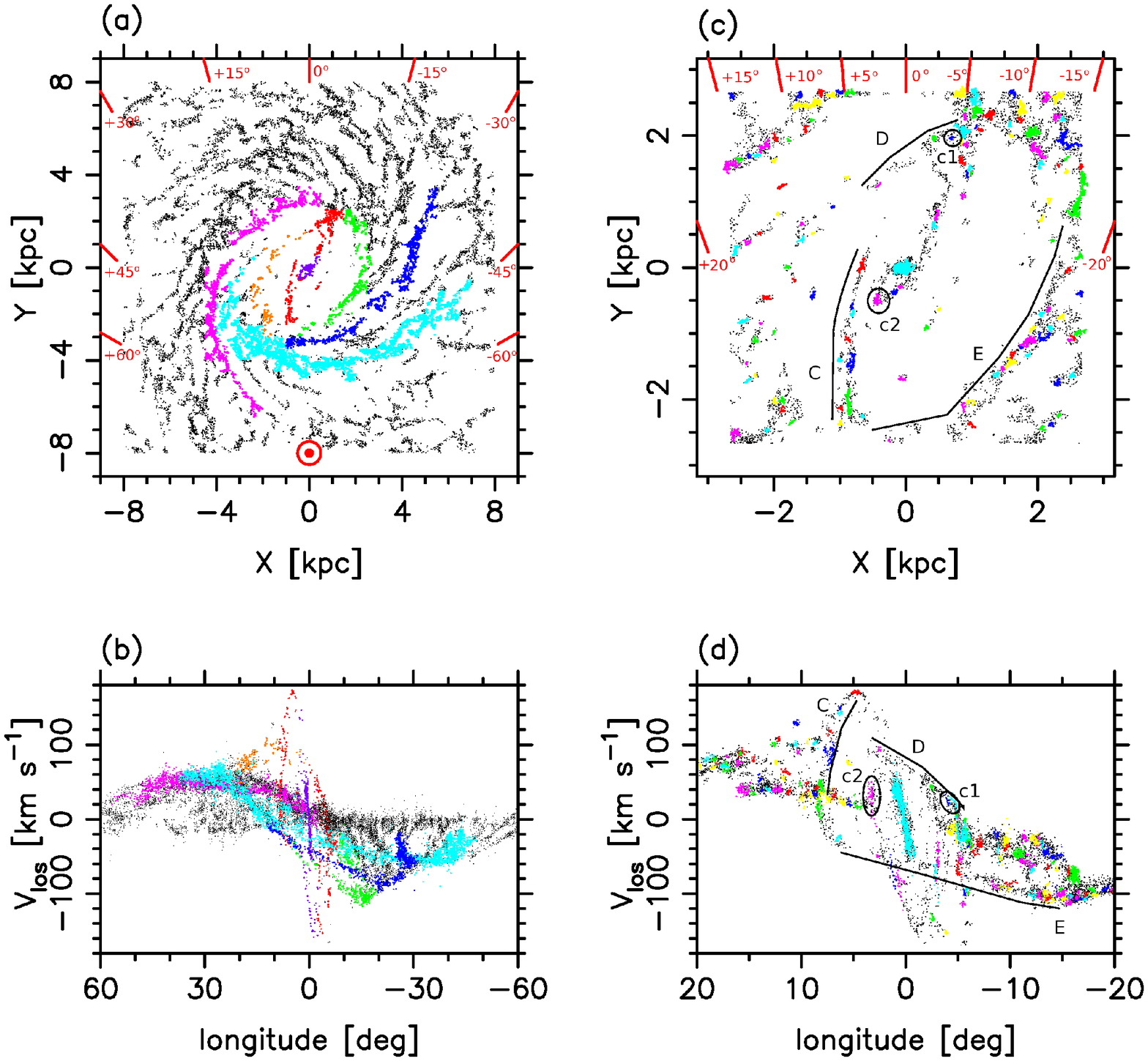}
\end{center}
\caption{
    (Left panels) Correspondence between the gaseous spiral arms in the
    $x-y$ plane and the $l-v$ features for the cold gas in the inner
    galaxy at $t=1.24~{\rm Gyr}$. (a): Spacial distribution on the $x-y$
    plane.  (b): $l-v$ diagram. Each SPH particle is plotted as a dot in 
    spite of its distance. The observer is located at 
    $(R_0,\phi_{\rm b})=(8~{\rm kpc}, 25^\circ)$ presented by the red
    mark point in the top panel, and has a pure circular motion.
    (Right panels) Same as the left panels, but colored for gas clumps.
    Solid curves labeled by ``C'', ``D'', and ``E'' correspond to them
    in Figure \ref{fig:pvschematic}. Open circles (c1 and c2) are
    examples of gas clumps around their apocenters of orbits. 
}
\label{fig:faceonlv}
\end{figure*}

In Figures \ref{fig:faceonlv}c and \ref{fig:faceonlv}d, we marked gas
clumps ($>10^5~\Mo$) with the same color between the face-on map and
$l-v$ diagram.  We here defined the gas clumps by using a
Friend-of-Friend (FOF) method \citep{Davis+1985} with a linking length
of $30$ pc for the cold gas.  The large-scale peculiar features, such as
the ``Connecting arm''(C), ``3-kpc arm''(E) and ``135-km s$^{-1}$
arm''(D) in $l-v$ diagram are actually ensembles of dense clouds and
filaments.

It is known that the Bania's ``Clump 1'' and ``Clump 2'' are placed
around the negative longitude end of the 135 km s$^{-1}$ arm and at the
positive longitude side of the CMZ in observed CO $l-v$ diagram 
\citep{Bania1977,Bania+1986,StarkBania1986}.
We associate these clumps with clumps ``c1'' and ``c2'' in Figures
\ref{fig:faceonlv}c and 7d, respectively. 
Mass of the clump ``c1'' is $\sim 10^5~\Mo$, the clump ``c2'' is
$\sim 5\times 10^5~\Mo$. 
From time evolution of these clumps, we found that the clumps ``c2''
eventually collapsed into the galactic center.
Other gas clumps within a bar region tend to spiral into the galactic center
with losing their angular momenta around their orbital apocenters via 
collisions between them.  
\citet{Fux1999} inferred that the Bania's clump 2 and another vertical
feature near the clump could be correspond to the gas clumps which are
just going to cross the near-side offset axis (i.e. connecting arm). 
Contrary to this picture, we here suggest that vertical features in
observed CO $l-v$ diagrams near the CMZ are streams of gas clumps
spiraling into the galactic center.
As a consequence, our simulation suggests that gas is stochastically
supplied into the galactic central region by dense gas clumps.

\subsection{The Non-Stationarity of $l-v$ Features}
We find that the observed $l-v$ diagram is reproduced only at 
a specific time ($t \sim 1.24$ Gyr) and 
their features change on a time-scale of $\sim 100$ Myr, 
suggesting that the observed $l-v$ features are transient features.
The non-stationarity has been also found by \citet{Fux1999} and
\citet{Bissantz+2003} in their numerical models. 
\citet{Fux1999} and \citet{Bissantz+2003} attributed the
non-stationarity to wandering of a stellar bar around the center of the
mass, and the difference of the pattern speeds between the stellar bar 
and spiral arms, respectively.  
In fact, our model shows that the stellar spiral arms rotate slowly than 
the bar (Figure \ref{fig:angfreq}b), 
suggesting that decoupling between the stellar bar and spirals may cause
the non-stationarity.

Recently, \citet{Baba+2009} analyzed kinematics of star forming regions
in the same numerical model discussed in this paper with observed proper 
motions of maser sources in the MW, 
and suggested that the Galactic stellar spiral arms should be transient, 
recurrently formed structures rather than the ``stationary'' density waves 
proposed in \citet{LinShu1964} (for a review, see \cite{BertinLin1996}).  
This transient nature of stellar spirals is also supported by previous 
$N$-body simulations of stellar disks without the ISM 
\citep{SellwoodCarlberg1984,Sellwood2000,Sellwood2010,Fujii+2010}. 
We infer that dynamic nature of stellar spiral arms themselves could
contribute the non-stationarity of the $l-v$ diagram. 
We will quantitatively investigate a driving mechanism of 
the non-stationarity in our following paper.

\section{Conclusions}
By using high-resolution, $N$-body+hydrodynamical simulation in which
the multi-phase ISM, star-formation, and SN feedback were
self-consistently taken into account, we qualitatively reproduced not
only large-scale structures of the H\emissiontype{I} and CO $l-v$
diagrams such as the terminal velocity tangent points and the coherent
features, but also clumpy structures.  Previous studies with numerical
simulations on $l-v$ diagrams did not reproduce these clumpy structures.
When we adopt a model galaxy whose velocity is similar to
the Milky Way galaxy with the same numerical method, we can advance our
argument more qualitatively.  We will show the results in the near
future.

\bigskip 
The authors are grateful to the anonymous referee for his/her valuable
comments. 
We would like to thank James Binney for his constructive comments for
our previous paper, by which led us to analyze our results using the
$l-v$ diagram.  We also thank William Robert Priestley for careful
reading the manuscript.  Calculations and visualization were performed
by Cray XT-4 in Center for Computational Astrophysics, National
Astronomical Observatory of Japan.  This project is supported by the
Molecular-Based New Computational Science Program, NINS. TRS is
financially supported by a Research Fellowship from the Japan Society
for the Promotion of Science for Young Scientists.

\bibliographystyle{apj}
\bibliography{ms}

\end{document}